\documentstyle{mn}

\def\beq{\begin{equation}}
\def\eeq{\end{equation}}
\def\bey{\begin{eqnarray}}
\def\eey{\end{eqnarray}}
\def\beqarray{\begin{eqnarray}}
\def\eeqarray{\end{eqnarray}}
\def\rE{r_{\rm E}}
\def\ME{M_{\rm E}}

\def\Dd{D_{\rm d}}
\def\zd{z_{\rm d}}
\def\Ds{D_{\rm s}}
\def\Dds{D_{\rm ds}}

\def\d{ {\rm d}}
\def\pibytwo{{\frac{\pi}{2}}}
\def\ffrac#1#2{{\textstyle\frac{#1}{#2}}}

\def\thetacusp{{\theta_{\rm cusp}}}
\def\nwea{{A}}
\def\nweb{{B}}
\def\spose#1{\hbox to 0pt{#1\hss}}
\def\lta{\mathrel{\spose{\lower 3pt\hbox{$\sim$}}
    \raise 2.0pt\hbox{$<$}}}
\def\gta{\mathrel{\spose{\lower 3pt\hbox{$\sim$}}
    \raise 2.0pt\hbox{$>$}}}

\input epsf
\title[Are there sextuplet and octuplet image systems?]        
{ARE THERE SEXTUPLET AND OCTUPLET IMAGE SYSTEMS?}

\author[N. Wyn Evans and Hans J. Witt]
{N. Wyn Evans$^1$ and Hans J. Witt$^2$ \\
$^1$ Theoretical Physics, 1 Keble Rd, Oxford, OX1 3NP \\
$^2$ University of Manchester, Jodrell Bank Observatory, Macclesfield,
Cheshire, SK11 9DL}

\pubyear{2001}

\begin{document}
\maketitle
\begin{abstract}
We study gravitational lensing by the family of scale-free galaxies
with flat rotation curves. The models are defined by a shape function,
which prescribes the radius of the isophote as a function of position
angle from the major axis.  The critical curves are analytic, while
the caustic network is reducible to a simple quadrature.  The cusps
are always located at the turning points of the shape function.  We
show that the models with exactly elliptic isophotes never admit
butterfly or swallowtail cusps and so have at most 4 (or 5)
images. Higher order imaging is brought about by deviations of the
isophotes from pure ellipses -- such as pointedness caused by embedded
disks or boxiness caused by recent merging.  The criteria for the
onset of sextuple and octuple imaging can be calculated analytically
in terms of the ellipticity $\epsilon$ and the fourth-order Fourier
coefficients ($a_4$ and $b_4$) used by observers to parametrise the
isophote shapes.  The 6 or 8 images are arranged roughly in a circle,
which appears as an incomplete Einstein ring if inadequately resolved.
Using data on the shapes of elliptical galaxies and merger remnants,
we estimate that $\sim 1 \%$ of all multiply imaged quasars may be
sextuplet systems or higher. Forthcoming satellites like the Global
Astrometric Interferometer for Astrophysics (GAIA) will provide
datasets of $\sim 4000$ multiply imaged systems, and so $\sim 40$ will
show sextuple imaging or higher.
\end{abstract}


\section{INTRODUCTION}

Almost all the known sixty or so gravitational lens systems are 2 or 4
image configurations (see Pospieszalka et al. 1999 for details of the
gravitational lensing database which maintains a list of
candidates). There are two systems in which the presence of a weak,
central image has been claimed, namely APM08279+5255 (Ibata et
al. 1999) and MG1131+0456 (Chen \& Hewitt 1993), but these detections
are controversial.  There are only two lens systems which are
definitely established to have more than 4 images of a single
source. The first is CL0024+1654 which comprises 8 images of a single
blue galaxy lensed by a foreground cluster of galaxies (Colley, Tyson
\& Turner 1996; Tyson, Kochanski \& Dell'Antonio 1998).  The second is
B1359+154, which consists of 6 images of a single radio source (Rusin
et al 2001). Here, the lens is believed to be a compact group of
galaxies.

For non-singular lenses, it is well-known that the total number of
images is odd and that the number of even parity images exceeds the
number of odd parity images by one (e.g., Burke 1981; Schneider,
Ehlers \& Falco 1992, chap. 5). When the source lies within the
central or tangential caustic, such lenses typically give rise to 5
images. Four of these lie roughly on an Einstein ring, while the fifth
is highly de-magnified and located near the centre of the lensing
galaxy.  For cusped lenses, the convergence (or surface density in
units of the critical density) $\kappa$ behaves like
\begin{equation}
\lim_{r \rightarrow 0} \kappa(r) = O(r^{-\gamma}),
\end{equation}
and so diverges at the galaxy centre.  Evans \& Wilkinson (1998)
showed that there are 4 images if the central cusp is stronger than
isothermal ($\gamma >1$), and 5 images if the cusp is weaker than
isothermal ($\gamma <1$), again assuming the source lies within the
central caustic. Hence, the observed abundance of 4 image systems may
be caused either by strong cusps (in which case there are only 4
images) or by small core radii (in which case the fifth image is so
de-magnified as to be invisible against the much brighter lensing
galaxy).

The purpose of this paper is to ask whether sextuplet or octuplet
images may exist? If so, what properties of the lensing galaxy cause
such higher order imaging? Note that in this paper, by a slight abuse
of terminology, we use sextuple imaging to refer to systems with 6 or
7 images; a central seventh image may exist if the galaxy is
soft-centered but will in general not be discernible.  Similarly,
octuple imaging may refer to 8 or 9 images.  Systematic programs like
the Cosmic Lens All-Sky Survey or CLASS (e.g., Myers et al 1995, Myers
1999) are now providing a much larger dataset of multiply-imaged
quasars than the earlier serendipitous discoveries.  The future holds
out even brighter prospects, as the Next Generation Space Telescope
(NGST) and the Global Astrometric Interferometer for Astrophysics
(GAIA) will discover many thousands of such systems (e.g., Barkana \&
Loeb 2000; de Boer et al 2000). Hence, there are golden opportunities
for uncovering comparatively unusual image configurations.

Section 2 provides an introduction to the lensing properties of
scale-free galaxies with flat rotation curves.~\footnote{Such models
are often called ``isothermal'' in the lensing literature. This is an
unfortunate nomenclature, as the velocity distributions are not in general
isothermal. We prefer to use the more cumbersome term ``scale free
galaxies with flat rotation curves'' to describe this family of
models. The circular velocity curve of the spherical and axisymmetric
models is everywhere flat. For the triaxial models, the circular
velocity curve is not strictly speaking defined. Nonetheless, the
structure of velocity space (and hence the elliptic closed orbit in
the principal plane on which any cold gas lies) is the same
independent of position. Hence, these models too have everywhere flat
rotation curves.} We show that the caustic network, together with the
locations of the cusps, can be worked out exactly. Section 3 considers
two simple and very familiar models, namely those with elliptic
isophotes and elliptic equipotentials (e.g., Kassiola \& Kovner
1993). Such models, if acting as isolated lenses, are unable to
produce higher order imaging. Section 4 identifies the physical
causes of higher order imaging in terms of deviations of the isophotal
shapes of the lensing galaxy from pure ellipses. Simple models are
provided for lenses with boxy, disky and squarish isophotes and the
onsets of sextuple and octuple imaging are calculated
analytically. Finally, Section 5 concludes with a discussion of the
frequency with which such higher order image systems occur, together
with an assessment of forthcoming observational opportunities.

\section{SCALE-FREE MODELS WITH FLAT ROTATION CURVES}

\subsection{Mass and Potential}

Scale-free galaxy models with flat rotation curves are widely used in
galactic astronomy and dynamics (e.g., Toomre 1982; Evans 1993; Evans,
Carollo \& de Zeeuw 2000).  The isophotes of a scale-free galaxy have
the same shape at every radius. The isophotes are completely described
by a {\em shape function} $G(\theta)$, which depends only on the
position angle $\theta$ with respect to the major axis.  In such
galaxies, the convergence $\kappa$ and deflection potential $\phi$
satisfy
\begin{equation}
\kappa = {G(\theta) \over 2r}, \qquad\qquad \phi = rF(\theta),
\label{eq:scalefree}
\end{equation}
where ($r,\theta$) are familiar polar coordinates in the lens plane.
The surface density falls off along any ray like the reciprocal of
distance and so the three-dimensional density falls off like the
inverse square of distance. The models are projections of axisymmetric
and triaxial generalisations of the familiar isothermal sphere.  As
$\nabla^2 \phi = 2 \kappa$ (see e.g., Schneider et al. 1992), it is
straightforward to establish that
\begin{equation}
G(\theta) = F(\theta) + F^{\prime\prime} (\theta).
\label{eq:ode}
\end{equation}
Given an arbitrary shape function $G(\theta)$, it is possible to solve
(\ref{eq:ode}) by the method of variation of the parameters (e.g.,
Bronshtein \& Semendyayev 1998, section 3.3.1.3.4) to obtain
\begin{equation}
F(\theta) = \sin \theta \int_0^\theta G(\vartheta) \cos \vartheta\,
\d\vartheta - \cos \theta \int_{\pibytwo}^\theta G(\vartheta) \sin
\vartheta\,\d\vartheta.
\label{eq:crux}
\end{equation}
The deflection angle has components
\begin{equation}
\phi_x = -\int_{\pibytwo}^\theta G(\vartheta) \sin
\vartheta\,\d\vartheta, \qquad
\phi_y = \int_0^\theta G(\vartheta) \cos \vartheta\,\d\vartheta.
\end{equation}
This equation relates the lensing properties of the galaxy directly to
the shape of the isophotes $G(\theta)$.

\subsection{Critical Curves and Caustics}

For scale-free models with flat rotation curves, the magnification is
simply related to the convergence via (e.g., Witt, Mao \& Keeton 2000)
\begin{equation}
\det J = 1 - 2 \kappa.
\end{equation}
The critical curves are given by the vanishing of the Jacobian
and so they have the polar equation
\begin{equation}
\label{eq:critcurve}
r = G(\theta) = F(\theta) + F^{\prime\prime} (\theta).
\end{equation}
Let the Cartesian coordinates of the caustic be $(\xi, \eta$). Then,
the caustic is given through the lens equation as
\begin{eqnarray}
\xi &=& F^{\prime\prime} (\theta) \cos\theta + F^\prime (\theta) \sin
\theta, \nonumber \\ 
\eta &=& F^{\prime\prime} (\theta) \sin \theta - F^\prime (\theta) \cos \theta.
\label{eq:caustic}
\end{eqnarray}
If we introduce polar coordinates ($s, \alpha$) in the source plane,
then the equations (\ref{eq:caustic}) become
\begin{equation}
s \cos (\theta - \alpha) = F^{\prime\prime}(\theta),\qquad
s \sin (\theta - \alpha) = F^{\prime}(\theta).
\end{equation}
By substituting (\ref{eq:crux}) into (\ref{eq:caustic}), we obtain the
result
\begin{eqnarray}
\xi &=& G(\theta) \cos \theta  + \int_{\pibytwo}^\theta G(\vartheta)
\sin \vartheta\, \d \vartheta, \nonumber \\
\eta &=& G(\theta) \sin \theta - \int_0^\theta G(\vartheta) \cos
\vartheta \, \d \vartheta.
\end{eqnarray}
Although this equation is simple to derive, it is rather remarkable.
For scale-free galaxies with flat rotation curves, the caustics can be
computed by a simple quadrature for {\em any isophotal shape}.
Generally speaking, in gravitational lensing it is very hard to
establish the maximum number of images that can be generated by a
given lens. Often, numerical searches are the only way of tackling
such questions and they risk missing small but important domains of
parameter space. By contrast, here the caustic network is directly
available to us and the onset of sextuple and octuple imaging can be
found exactly.

\subsection{The Role of Cusps}

The appearance of swallowtails or butterflies usually leads to areas
in the source plane where 6 or more images are possible. Knowledge
about the position of cusps is therefore very important. Since the
caustic is given by the parametric representation (\ref{eq:caustic}),
it is straightforward to derive the cusp locations. At a cusp, the
caustic stops and reverses direction, so that the derivative of the
caustic coordinates with respect to the parameter must vanish:
\begin{equation}
\label{eq:cusp}
\frac{d\xi}{d \theta}  = G^\prime(\theta) \cos\theta = 0 , \qquad
\frac{d\eta}{d \theta} = G^\prime(\theta) \sin\theta = 0. 
\end{equation}
Both equation can only vanish simultaneously if the common factor
$G^\prime(\theta)$ vanishes.  Therefore, the location of the cusps
$\thetacusp$ is simply given by the equation $G^\prime(\thetacusp) =
0$. This fact is remarkable since usually the equation for the cusps
is very complicated. The cusps can occur only at the turning points of
the shape function, so they are related to the properties of the
isophotes in a straightforward way for these models.

In models without higher order cusps -- such as swallowtails and
butterflies -- the caustic only has 4 simple cusps or folds. If a
galaxy model permits the occurrence of swallowtails or butterflies,
then it can show sextuple or octuple imaging.  The onset of
swallowtails happens when the second derivative of the shape function
vanishes as well, $G^\prime(\thetacusp) = G^{\prime\prime}(\thetacusp)
= 0$ (see Schneider et al.  1992; Keeton, Mao \& Witt 2000) so the
first non-zero derivative is the third. Similarly, the onset of
butterflies happens when all of the first three derivatives of the
shape function vanish, $G^\prime(\thetacusp) =
G^{\prime\prime}(\thetacusp) = G^{\prime\prime\prime}(\thetacusp) =
0$, so the first non-zero derivative is the fourth.

The zeros for the cusps $\thetacusp$ are related to the length
of the caustic. For the length of the caustic, we may write
\begin{eqnarray}
l_{\rm caustic} & = & \int_0^{2\pi} \sqrt{(\xi^\prime(\theta))^2 
+ (\eta^\prime(\theta))^2} d\theta = 
\int_0^{2\pi} \vert G^\prime(\theta) \vert d\theta \nonumber \\
& = & 2 \sum_{i=1}^{n_{\rm cusp}/2} \vert G(\theta_{{\rm cusp},2i})
- G(\theta_{{\rm cusp},2i-1}) \vert.
\end{eqnarray}
Finally, the area under the caustic can also be given 
\begin{eqnarray} 
A_{\rm caustic} &=&\int_0^{2\pi} \xi(\theta) \eta^\prime(\theta)
d\theta \nonumber \\
&=& \int_0^{2\pi} d\theta\int_0^\theta d\psi G^\prime( \theta) G^\prime
(\psi) \cos \theta \sin \psi.
\end{eqnarray}
This holds good provided the caustic does not self-intersect.
\begin{figure}
\epsfysize=7.5cm \centerline{\epsfbox{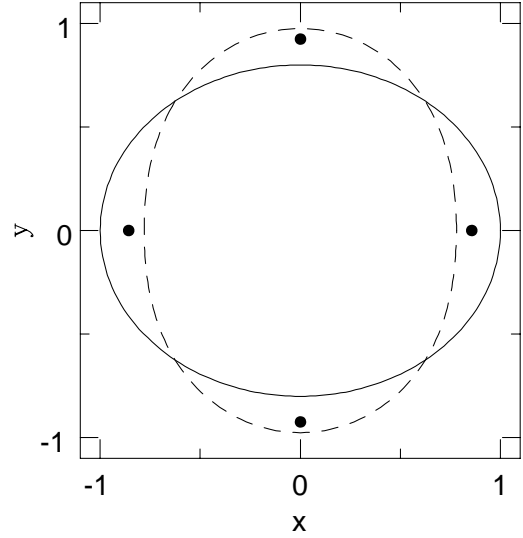}}
\caption{Critical curve (solid line) and reciprocal critical curve [as
defined in eq.~(\ref{eq:defrcc})] (dashed line) for an elliptical
density distribution ($q=0.8$) is shown. In both cases the mass inside
the curve is proportional to the area, i.e. $A = \pi M$.  The filled
circles show the positions of the images when the source is located at
the origin.}
\label{fig:area}
\end{figure}
\begin{figure}
\epsfysize=7.5cm \centerline{\epsfbox{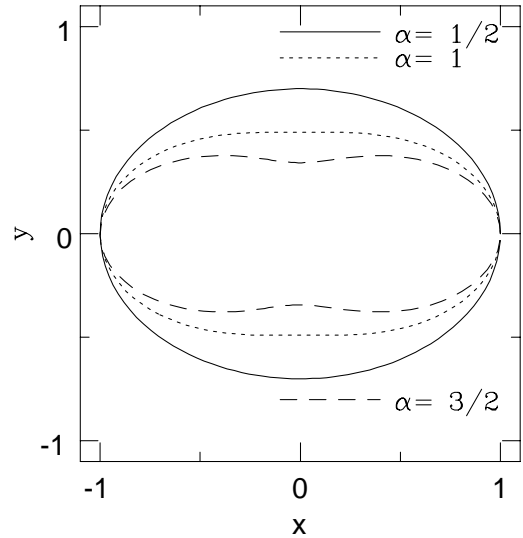}}
\caption{The isophotes of the family of models with shape function
(\ref{eq:genshapefunc}) with $q =0.75$. Note that $\alpha = 0.5$
corresponds to elliptical isophotes, while $\alpha = 1.5$ corresponds
to elliptical equipotentials. All these lenses are unable to show
higher order imaging.}
\label{fig:iso}
\end{figure}
\begin{figure*}
\epsfysize=12cm \centerline{\epsfbox{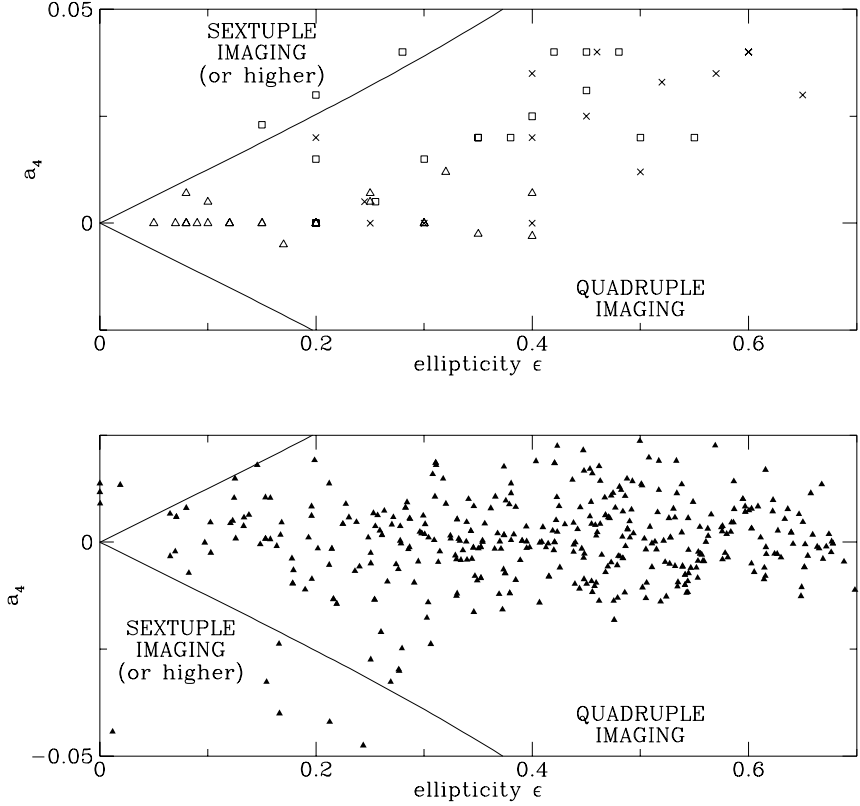}}
\caption{The onset of sextuple imaging as a function of the
ellipticity of the isophotes $\epsilon$ and the parameter $a_4$ used
to measure boxy and disky deviations. Upper panel additionally shows
data from Saglia et al. (1993); triangles are elliptical galaxies,
crosses are S0s and squares are SB0s in the Coma cluster. Lower panel
additionally shows data from simulations of merger remnants by Heyl et
al. (1994).}
\label{fig:onset}
\end{figure*}

\subsection{Area and Mass}

For the spherical case ($G(\theta) = {\rm const}$), it is well known
that the mass inside the Einstein ring is directly related to the
Einstein radius by $\ME = \rE^2$ (cf. Schneider et al. 1992).  The
area enclosed by the Einstein ring (i.e. the area inside the critical
curve) is related to the mass by $\pi M_{\rm crit. curve} = A_{\rm
crit. curve}$.  From an observational perspective, the relation
becomes $\pi M_{\rm crit. curve} = \Sigma_{\rm crit} \Dd^2 A_{\rm
crit. curve}$, where the area is measured in arcsec$^2$ and the
critical surface mass density is given by
\begin{equation}
\Sigma_{\rm crit} = \frac{c^2 \Ds}{4\pi G \Dd \Dds}.
\label{eq:critdens}
\end{equation}
Here, $\Dd, \Ds$ and $\Dds$ are the distance to the deflector, the
distance to the source and the distance from the deflector to the
source respectively.  Given the redshift of the lens and the source,
this (\ref{eq:critdens}) enables the projected mass within the
Einstein ring to be calculated for a spherical lens. In general, these
useful relations do not remain valid for arbitrary, non-spherical
lenses.

However, we now show that they do remain true for any scale-free
potential obeying eq.~(\ref{eq:scalefree}).  The mass within the
critical curve is
\begin{eqnarray}
M_{\rm crit. curve} &=&  \frac{1}{\pi}
\int_0^{2\pi} \int_0^{R(\theta)} \kappa(r,\theta) r dr d\theta
\nonumber \\
&=& \frac{1}{2\pi}  \int_0^{2\pi} G^2(\theta) d\theta.
\end{eqnarray}
The area inside the critical curve is
\begin{equation}
A_{\rm crit.\, curve} = 
\int_0^{2\pi} \int_0^{R(\theta)} r dr d\theta
= \frac{1}{2}  \int_0^{2\pi} G^2(\theta) d\theta,
\end{equation}
where we have used eq.~(\ref{eq:critcurve}) for the dependency of the
critical curve on the angle. So, we obtain again $\pi M_{\rm crit} =
A_{\rm crit}$.

Many authors have suggested that the mass inside the ellipsoidal ring
formed by the images of a quadruple lens is approximately constant,
almost independent of the lens model; for example, the models of
Kochanek (1991), Rix, Schneider \& Bahcall (1994), Witt, Mao \&
Schechter (1995), Chae, Turnshek, \& Khersonsky (1998) and Hunter \&
Evans (2001) for the Einstein Cross (Q2237+0305) are very different
and predict different flux ratios for the 4 images, but they all agree
on the projected mass within the central $0.9^{\prime\prime}$ enclosed
by the images.  We want to shed some light on this issue!  Let us
define {\it the reciprocal critical curve} as
\begin{equation}
R(\theta) = c/r_{\rm crit.\, curve} = c / G(\theta).
\label{eq:defrcc}
\end{equation}
The mass inside the reciprocal critical curve is
\begin{equation}
M_{\rm recip.\,curve} = \frac{1}{\pi} 
\int_0^{2\pi} \int_0^{R(\theta)} \kappa(r,\theta) r dr d\theta = c,
\end{equation}
and it too must remain constant. Let us choose the constant $c$ in
such a way that the area inside the reciprocal critical curve $A_{\rm
recip.\, curve} / \pi =c$ is proportional to the mass,
obtaining
\begin{equation}
c = {\displaystyle 2 \pi \over \displaystyle \int_0^{2\pi} G^{-2}
(\theta) d\theta}.
\end{equation}
As shown in Figure~\ref{fig:area}, we can imagine that the images are
confined by the critical curve and the reciprocal critical curve.  If
the source position is closer to the caustic, then the images are
located closer to the critical curve. In other cases, the images may
be closer to the reciprocal critical curve.  The image positions
follow (approximately) the critical curve or the reciprocal critical
curve. The area within the critical and inverse critical curves is
directly proportional to enclosed mass. Hence, to a good
approximation, the area enclosed by the four images is as well.
(Small deviations might arise due to the interpolation curve
$R(\theta)$ chosen to connect the images of the quadruplet).
\begin{figure}
\epsfysize=10cm \centerline{\epsfbox{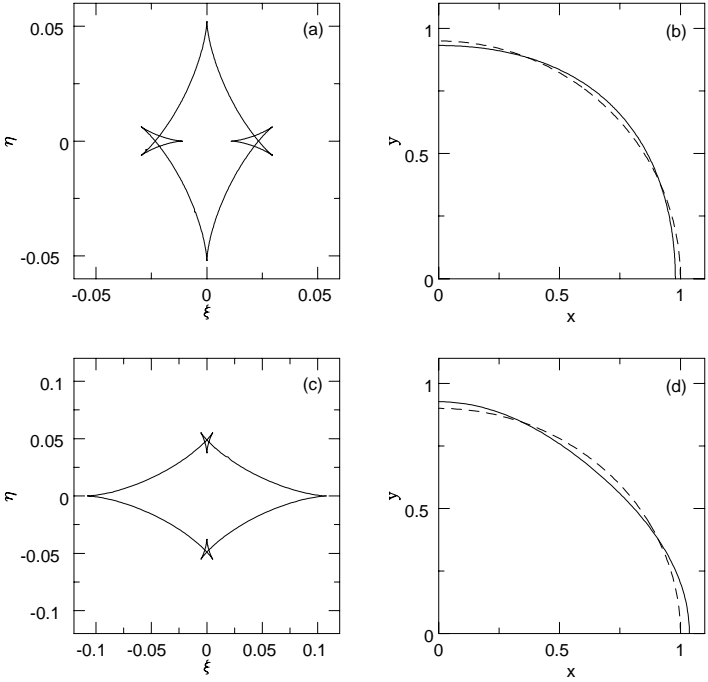}}
\caption{The caustic network and isophote shape for elliptical galaxy
models with (a,b) boxy and (c,d) disky isophotes. Butterfly
caustics appear on respectively the major and minor axes projected
onto the source plane. A pure elliptical isophote is shown for
comparison in dotted lines, so that the boxy and disky isophotal 
deviations can be discerned. [Panels (a,b) are for a model with
$\epsilon=0.05$ and $a_4 = -0.02$, panels (c,d) are for $\epsilon = 0.1$ and
$a_4 = 0.03$].}
\label{fig:cuspone}
\end{figure}
\begin{figure}
\epsfysize=5cm \centerline{\epsfbox{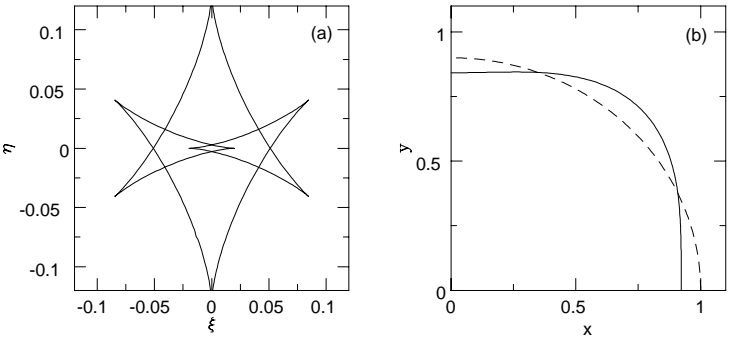}}
\caption{As the distortion increases, the butterfly cusps merge.
If the source lies in the region where the two caustics overlap, then
eightfold imaging occurs. [This shows a model with $\epsilon=0.1$ and
$a_4 = -0.075$].}
\label{fig:cusptwo}
\end{figure}

\section{Simple Models}

\subsection{Elliptic Isophotes and Equipotentials}

There are two very familiar models that fall into the class of
scale-free galaxies with flat rotation curves.  The first has elliptic
isophotes, so that the shape function is
\begin{eqnarray}
G(\theta) &\propto& \left(\ffrac{1}{2}(1+q^{-2}) +
\ffrac{1}{2}(1-q^{-2}) \cos 2\theta \right)^{-\frac{1}{2}},\nonumber
\\ &\propto& \left( \cos^2 \theta + q^{-2} \sin^2 \theta
\right)^{-\frac{1}{2}}.
\label{eq:ellipdens}
\end{eqnarray}
It is easy to find the deflection potential as
\begin{equation}
\phi = x\,\phi_x + y\,\phi_y,
\end{equation}
with
\begin{eqnarray}
\phi_{,x} &=& -\int_{\pibytwo}^\theta G(\vartheta) \sin \vartheta \,
\d\vartheta \nonumber \\
&=& {q \over \sqrt{1-q^2}}
\,\tan ^{-1}\left(\sqrt{1\!-\!q^2 \over q^2 \cos^2\theta\!+\! \sin^2\theta}\,
\cos\theta\right)\,, \nonumber\\
\phi_{,y} &=&  \int_{0}^\theta G(\vartheta) \cos \vartheta \,
d\vartheta \nonumber \\ &=& {q \over \sqrt{1-q^2}}
\,\tanh^{-1}\left(\sqrt{1\!-\!q^2 \over q^2 \cos^2\theta\!+\!\sin^2\theta}\,
\sin\theta\right)\,.
\end{eqnarray}
This result has been given before (see e.g., Kassiola \& Kovner 1993;
Kormann, Schneider, Bartelmann 1994). By explicit construction, Keeton
et al. (2000) showed that an elliptical density distribution plus
external shear can provide 6 or 8 images configurations. This happens
when the external shear is strong and aligned along the minor axis of
the lensing galaxy.

The second has elliptic equipotentials, so that 
\begin{eqnarray}
F(\theta) &\propto& \sqrt{\ffrac{1}{2}(1+q^{-2}) + \ffrac{1}{2}(1-q^{-2}) \cos
2\theta} \nonumber \\
&\propto& \sqrt{ \cos^2 \theta + q^{-2} \sin^2 \theta }.
\end{eqnarray}
This model was introduced into lensing by Blandford \& Kochanek (1987)
and subsequently studied by others (e.g., Witt 1996; Witt \& Mao 1997,
2000; Hunter \& Evans 2001).  The three dimensional analogues of these
models are well-known in galactic dynamics as power-law galaxies
(Evans 1993, 1994). It is straightforward to find the shape function
as
\begin{equation}
G(\theta) \propto q^{-2}\left(\ffrac{1}{2}(1+q^{-2}) +
\ffrac{1}{2}(1-q^{-2}) \cos 2\theta \right)^{-\frac{3}{2}}.
\end{equation}
On geometric grounds, it seems likely that an elliptically stratified
potential plus external shear cannot have more than 4 (or 5) images.
The reason for this is the confinement of the image positions. As
found by Witt (1996) and Witt \& Mao (1997), the 4 (or 5) images of an
elliptical potential are constrained to lie on a hyperbola.  If any
new image pairs appear, they must also lie on this
hyperbola. Physically, we expect that each new image pair is located
close to the Einstein ring. Since a hyperbola can have at most only 4
intersection points with a ring-like curve, it is intuitively clear
that an elliptical potential plus shear may not have more than 4
images (or 5 images if there is additionally one close to the centre
of the lensing galaxy). An analytical proof of the maximum number of
images for power-law galaxies with elliptic equipotentials is provided
in Appendix~A of Hunter \& Evans (2001).
\begin{figure}
\epsfxsize=9cm \epsfysize = 6cm \centerline{\epsfbox{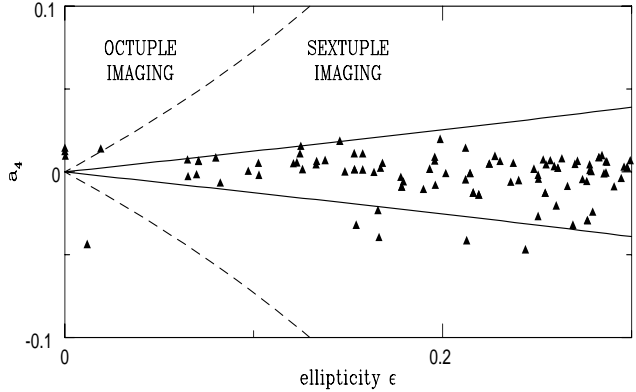}}
\caption{The onset of octuple imaging as a function of the
ellipticity of the isophotes $\epsilon$ and the parameter $a_4$ used
to measure boxy and disky deviations. Data from simulations of merger
remnants by Heyl et al. (1994) are overplotted.} 
\label{fig:onset8}
\end{figure}

\subsection{Number of Images}

Both these simple models are members of a more general class that have
a shape function $G(\theta)$ of the form:
\begin{equation}
G(\theta) \propto
\left(\ffrac{1}{2}(1+q^{-2}) + \ffrac{1}{2}(1-q^{-2}) \cos
2\theta \right)^{-\alpha},
\label{eq:genshapefunc}
\end{equation}
where $\alpha$ is constant. Figure~\ref{fig:iso} shows isophotes for
three members of the family, namely $\alpha = 1/2$ (elliptic
isophotes), $\alpha =1$ and $\alpha = 3/2$ (elliptic equipotentials).
For the same $q$, the models become flatter with increasing $\alpha$.
The isophotes can become ``peanut-shaped''; this is valuable for
modelling the peanut-shaped bulges of spiral galaxies (see e.g., Shaw
1987, Bureau \& Freeman 1999).

The positions of the cusps are found by solving $G^\prime(\theta) =0$
and so
\begin{equation}
\thetacusp = {n\pi \over 2}, \qquad\qquad n =0,1,2,3.
\end{equation}
It is straightforward to establish
\begin{eqnarray}
G^{\prime\prime}(\thetacusp) \propto (1\!-\! q^{-2})
\left[1\!+\!q^{-2}\!+\! (-1)^n(1\!-\!q^{-2})
\right]^{-\alpha\!-\!1}.
\end{eqnarray}
For flattened models ($q \neq 1$), this never vanishes and so higher
order cusps like swallowtails never occur. This provides a simple
proof that scale-free galaxies with flat rotation curves and elliptic
isophotes or equipotentials have at most 4 (or 5) images depending on
the density singularity at the centre. Of course, the proof is
slightly more general, holding good for all models with shape
functions of form eq~(\ref{eq:genshapefunc}).

\section{Isophotal Deviations}

The isophotes of elliptical galaxies are well approximated by pure
ellipses. Observers typically measure deviations from ellipses of the
order of a few percent. This might betray the presence of an embedded
disk or a central bar. Boxy or irregular isophotes may be caused by
recent merging and interaction.  Typically the difference between the
isophote $r_{\rm i}$ and a reference ellipse $r_{\rm e}$ is expanded
as a Fourier series (e.g., Binney \& Merrifield 1998)
\begin{equation}
r_{\rm i} - r_{\rm e} = \frac{a_0}{2} + 
\sum_{n=1}^\infty a_n \cos n \theta + b_n \sin n \theta,
\end{equation}
Provided the reference ellipse has been well-chosen, the Fourier
coefficients are all small. The largest non-vanishing components are
usually $a_4$ (which is a measure of boxiness and diskiness) and $b_4$
(which is a measure of skewness and squareness). The $a_3$ and $b_3$
coefficients are significant only in elliptical galaxies with dust
lanes. Observers normalise the Fourier coefficients to the semimajor
axis of the reference ellipse. Henceforth, we always use normalised
Fourier coefficients, so that our models are directly comparable to
the data.

\begin{figure*}
\epsfysize=10cm \centerline{\epsfbox{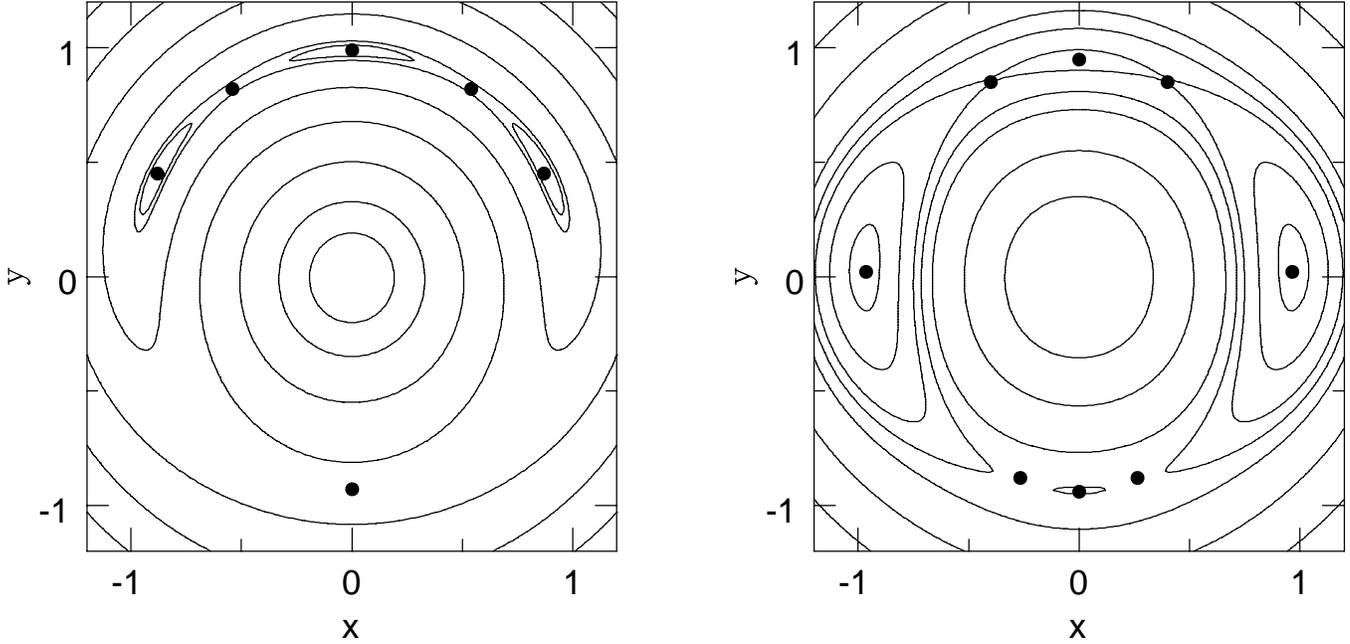}}
\caption{The topology of the Fermat surfaces for instances of (left
panel) sextuple and (right panel) octuple imaging. The even parity
images are located at the minima of the Fermat surface, the odd
parity images at the saddle points. Image locations are marked with
filled circles. [The left panel corresponds to the model shown in
Figure~\ref{fig:cuspone}(a) with source position $\xi = 0.02, \eta
=0.0$. The right panel corresponds to the model shown in
Figure~\ref{fig:cusptwo}(a) with source position $\xi = 0.005, \eta
=0.0$].}
\label{fig:fermat}
\end{figure*}

\subsection{Boxiness and Diskiness}

Boxiness and diskiness has been of interest ever since Bender et
al. (1989) pointed out possible correlations of the $a_4$ parameter
with X-ray, radio and kinematic properties. This correlation remains
poorly understood, especially as the boxiness and diskiness is partly
a function of the inclination angle of the galaxy (Franx 1988).  The
flattest galaxies have the largest values for $|a_4|$. In fact,
galaxies with an ellipticity greater than $0.35$ almost always have
$|a_4| >0.005$ (Bender et al.  1989). Approximately one third of
elliptical galaxies show boxiness at the level $a_4 < -0.005$, one
third show diskiness at the level $a_4 > 0.005$ and one third show
deviations of $|a_4| < 0.005$.  Values of $|a_4|$ much in excess of
$0.05$ are uncommon for normal ellipticals.

We study the shape function is
\begin{equation}
G(\theta) \propto \left[\ffrac{1}{2}(1+q^{-2}) +
\ffrac{1}{2}(1-q^{-2}) \cos 2\theta - \nwea \cos 4\theta) 
\right]^{-\frac{1}{2}}.
\end{equation}
When $\nwea$ vanishes, the model reduces to
(\ref{eq:ellipdens}). Otherwise, if $\nwea >0$, then the model has
disky isophotes, while if $\nwea <0$ then the model has boxy
isophotes. Observers typically record the isophote shape in terms of
the ellipticity $\epsilon = 1-q$ and the Fourier coefficient $a_4$.
In the limit of small deviations, $\nwea$ is related to the
observables through the equation
\begin{equation}
a_4 = \frac{\nwea}{32} (4-3\epsilon)^2 + O(\epsilon^4, \nwea^2). 
\end{equation}
These models can exhibit butterfly cusps at surprisingly small
deviations. The condition for butterflies is $G^\prime (\thetacusp) =
G^{\prime\prime} (\thetacusp) = G^{\prime\prime\prime} (\thetacusp)
=0$, so that the first non-vanishing derivative is the fourth. This is
satisfied at or beyond the critical value
\begin{equation}
|\nwea_{{\rm crit,}6}| = \frac{1}{8}(1 - q^{-2}).
\end{equation}
This condition can be re-written in terms of the observational
parameters $a_4$ and $\epsilon$ as
\begin{equation}
|a_4| \gta \frac{1}{256} {\epsilon (4 - 3\epsilon)^2 (2-\epsilon)
\over ( 1- \epsilon)^2}.
\label{eq:cruxa}
\end{equation}
This marks the onset of sextuple imaging. 

In Figure~\ref{fig:onset}, the boundary between quadruple and sextuple
imaging is shown in solid lines. In the upper panel, we also plot
the data of Saglia, Bender \& Dressler (1993) on a total of 58
elliptical galaxies, S0s and SB0s in the Coma cluster. In the lower
panel, we plot the data on isophotal shapes of 424 merger remnants
from simulations by Heyl, Hernquist \& Spergel (1994).  There are a
number of things to notice. First, very small values of $|a_4|$
suffice to produce sextuple imaging, if the galaxy is quite
round. However, nearly round galaxies tend to have very small
deviations from the pure elliptical form. Second, all the elliptical
galaxies in Saglia et al.'s (1993) sample lie in the quadruple imaging
domain. The possibilities for sextuple imaging improve for S0 and
especially SB0 galaxies. Saglia et al's sample contains only galaxies
for which a stable value of $a_4$ can be derived from the photometry
and does not include galaxies which exhibit isophote twisting or those
which show boxiness and diskiness in the same isophotal contours.  So,
it probably underestimates the frequency with which substantial
deviations from pure ellipses occur. Nonetheless, out of a total of 58
galaxies, 3 have the potential for sextuple imaging.  Third, the
merger remnants of Heyl et al. (1994) are typically boxy. This though
is a rough rule-of-thumb, as some can appear disky.  We see that 10
out of the total of 424 simulated merger remnants are irregular enough
to permit sextuple imaging or higher. Bearing in mind that most of the
lensing optical depth is in early-type galaxies, we reckon that the
fraction of lenses that have the ability to cause sextuple imaging or
higher is $\sim 3 \%$.

Figure~\ref{fig:cuspone} shows the caustic networks for two models. The
upper panels refer to a model which has ellipticity $\epsilon =0.05$
and boxy isophotes with $a_4 = -0.02$. The lower panels refer to a
model which has $\epsilon = 0.1$ and disky isophotes with $a_4 =
0.03$.  The locations of the cusps are given by the equations
\begin{equation}
\sin2\thetacusp = 0 \quad {\rm and} \quad \cos2\thetacusp = 
\frac{1-q^{-2}}{8\nwea}.
\end{equation}
This means that the 4 cusps remain on the major and minor axes with
increasing distortion $\nwea$, but 4 additional cusps appear as soon
as the critical value $\nwea_{{\rm crit},6}$ is exceeded.  Note the
butterfly cusps develop on the major axis in the boxy case, and on the
minor axis in the disky case.  If the source lies within the butterfly
caustic, then it is lensed into six images. 

The butterfly cusps touch at the onset of octuple
imaging.  Figure~\ref{fig:cusptwo} shows an example when the butterfly
cusps have merged.  A source in the overlapping region will be
lensed into eight images.  The isophote shape is now very rectangular,
but examples of such galaxies are known (see, for example, NGC 128 as
pictured in Burbidge \& Burbidge 1959).  The critical value of the
distortion $\nwea_{{\rm crit,}8}$ required for eightfold imaging is given
by solution of the implicit equation
\begin{eqnarray}
&&\int_0^\pibytwo {d \vartheta \cos \vartheta \over
[\ffrac{1}{2}(1\!+\!q^{-2})\!+\!\ffrac{1}{2}(1\!-\!q^{-2}) \cos
2\vartheta\!-\!|\nwea_{{\rm crit,}8}| \cos 4\vartheta]^{\frac{1}{2}}} 
\nonumber \\ &&= {1 \over \sqrt{q^2 - |\nwea_{{\rm crit,}8}|}}.
\end{eqnarray}
As the distortion is always small, we can linearise this
equation and solve it to obtain
\begin{equation}
|\nwea_{{\rm crit,}8}| \approx {2 (I_1-q)\over q^3 -I_2}.
\label{eq:cruxb}
\end{equation}
This marks the onset of octuple imaging. The integrals $I_1$ and
$I_2$ are
\begin{eqnarray}
I_1 &=& \int_0^{\pibytwo} { \cos \vartheta d \vartheta\over
[\ffrac{1}{2}(1+q^{-2}) + \ffrac{1}{2}(1-q^{-2}) \cos
2\vartheta ]^{\frac{1}{2}}}, \nonumber \\
&=& \frac{q}{\sqrt{1-q^2}} {\tanh}^{-1} \sqrt{1-q^2}, \nonumber \\
I_2 &=& \int_0^{\pibytwo} { \cos \vartheta \cos 4 \vartheta d \vartheta\over
[\ffrac{1}{2}(1+q^{-2}) + \ffrac{1}{2}(1-q^{-2}) \cos
2\vartheta ]^{\frac{3}{2}}}, \nonumber \\
&=& q[ q^4 + 10q^2 + 1 - 4q(2+q^2)I_1] / (1-q^2)^2.
\end{eqnarray}
They can be evaluated numerically easily enough.
Figure~\ref{fig:onset8} shows the distortion required for octuple
imaging in the plane of the observables. Overplotted is the simulation
data of Heyl et al. (1994), from which it is clear that 5 of the
merger remnants are sufficiently distorted to permit even eightfold
imaging.

Figure~\ref{fig:fermat} shows contours of the Fermat potential
\begin{equation}
\psi (x,y, \xi, \eta) = \ffrac{1}{2} | (x-\xi)^2 + (y - \eta)^2 |
- \phi(x,y),
\end{equation}
for typical 6 image and 8 image configurations.  Let us recall that
the Fermat potential is proportional to the time delay (e.g.,
Schneider et al. 1992).  The left panel shows the model of
Figure~\ref{fig:cuspone}(a) with the source placed within the
butterfly cusp so that it has 6 images (3 of even parity corresponding
to the minima of the Fermat surface, 3 of odd parity corresponding to
the saddle points). The right panel shows the model of
Figure~\ref{fig:cusptwo}(a) with the source placed in the region of
overlap of the butterfly cusps, so that there are now 8 images, 4 of
even parity and 4 of odd parity. Notice that all the images are at
roughly the same distance from the centre of the lensing galaxy. In
fact, the arrangement has obvious similarities with the deep {\it
Hubble Space Telescope} images of CL0024+1654, which comprises 8
images of a single blue galaxy lensed by a foreground cluster (Tyson,
Kochanski \& Dell'Antonio 1998), as well as the incomplete optical
Einstein ring 0047-2808, which is the image of a high redshift
early-type star-forming galaxy (Warren et al. 1999).

Scale-free models with flat rotations have another elegant and unusual
property. The relative time delay between the $i$th and $j$th image is
simply related to the radial positions $r_i$ and $r_j$ via (Witt et
al. 2000; see also Koopmans, de Bruyn \& Jackson 1998; Zhao \& Pronk
2001)
\begin{equation}
\Delta t_{i,j} = {\Dd\Ds \over 2c \Dds} (1 + \zd) (r_j^2 - r_i^2),
\end{equation}
where $\zd$ is the redshift of the deflector. Sextuple and octuple
image systems typically have all the images in a ring-like
configuration and so the relative time delays between any of the pairs
is never very large.
\begin{figure}
\epsfysize=10cm \centerline{\epsfbox{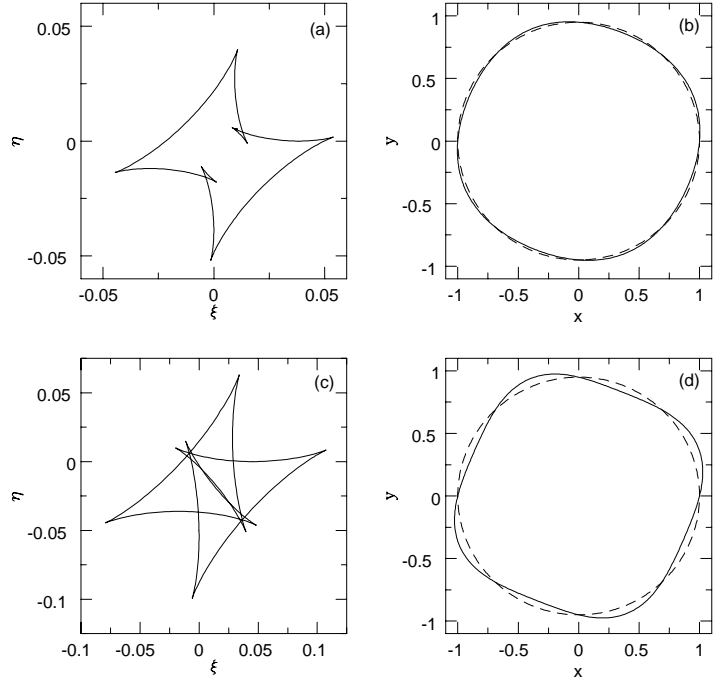}}
\caption{The caustic network and isophote shape for galaxies with the
shape function (\ref{eq:shapetwo}). Again, pure elliptical isophotes
are shown in dotted lines for comparison. Both models have
swallowtail cusps. [Panels (a,b) are for a model with $\epsilon =
0.05$ and $b_4 = 0.02$, panels (c,d) are for $\epsilon = 0.05$ and $b_4 =
0.06$].}
\label{fig:cuspthree}
\end{figure}

\begin{figure}
\epsfxsize=9cm \epsfysize = 6cm \centerline{\epsfbox{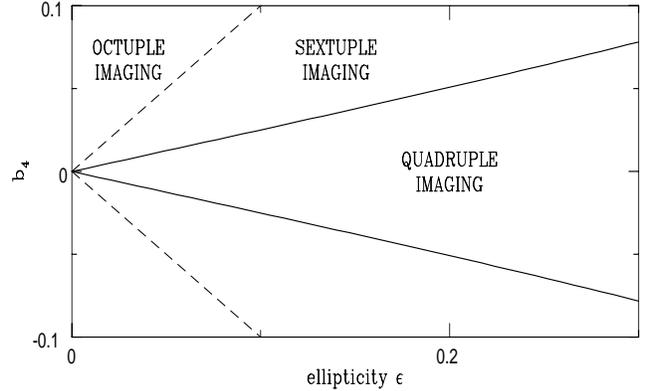}}
\caption{The onset of sextuple imaging as a function of the
ellipticity of the isophotes $\epsilon$ and the parameter $b_4$ used
to measure skewed squarishness.}
\label{fig:onsetb}
\end{figure}

\subsection{Skewed Squarishness}

Photometric studies of elliptical galaxies find that the other Fourier
coefficient that is typically non-zero is $b_4$ (e.g., Franx,
Illingworth \& Heckman 1989). To study its effect, we examine the
shape function 
\begin{equation}
G(\theta) \propto \left[\ffrac{1}{2}(1+q^{-2}) + \ffrac{1}{2}(1-q^{-2}) \cos
2\theta - \nweb \sin 4\theta) \right]^{\frac{1}{2}},
\label{eq:shapetwo}
\end{equation}
where without loss of generality $\nweb \ge 0$.  If the distortion
$\nweb$ is non-vanishing, then the isophotes become skewed and
squarish. This may be useful for modelling face-on bars in spiral
disks, rather like the lensing galaxy for Q2237+0305 (the Einstein
Cross). The observables are the ellipticity $\epsilon = 1-q$ and the
Fourier coefficient $b_4$, which is
\begin{equation}
b_4 = \frac{\nweb}{32} (4-3\epsilon)^2 + O(\epsilon^4, \nweb^2). 
\end{equation}
Unlike $a_4$, the sign of $b_4$ carries no physical significance and
can be inverted by a simple reflection. Observationally speaking, the
$b_4$ coefficient is usually, but not always, smaller than $a_4$ (see
e.g., Franx et al. 1989). The $b_4$ coefficient has received less
attention than $a_4$, as it does not appear to be correlated with
radio or X-ray properties. There are no extensive tables of data on
$b_4$ available in the literature. Nonetheless, individual galaxies
with $b_4$ as large as $0.05$ are known, such as NGC 1700 (Franx et
al. 1989).

The condition for swallowtails is $G^\prime (\thetacusp) =
G^{\prime\prime} (\thetacusp) =0$, so that the first non-vanishing
derivative is the third. This is satisfied whenever $\nweb$ exceeds
the critical value given
\begin{equation}
\nweb_{{\rm crit,}6} = \pm \frac{1}{4} (1-q^{-2}). 
\end{equation}
Again, it is useful to rewrite this in terms of the observables as
\begin{equation}
b_4 \approx \frac{1}{128} {\epsilon (4 - 3\epsilon)^2 (2-\epsilon)
\over ( 1- \epsilon)^2}.
\label{eq:cruxc}
\end{equation}
This marks the onset of sextuple imaging.

Figure~\ref{fig:cuspthree} shows the caustic network for a case when
the swallowtails have just developed ($\epsilon = 0.05$ and $b_4 =
0.02$), together with a case in which the swallowtails have overlapped
to create a domain in the source plane where eightfold imaging occurs
($\epsilon = 0.05$ and $b_4 = 0.06$).  The cusps occur at the
locations
\begin{equation}
\sin 2\theta = \frac{1-q^{-2}}{16\nweb} \pm 
\sqrt{\frac{(1-q^{-2})^2}{256\nweb^2} +\frac{1}{2}}.
\end{equation}
The cusps move further away from the major and minor axes for
increasing distortion $\nweb$. We obtain 4 additional cusps when the
swallowtails form as $\nweb$ reaches the critical value $\nweb_{{\rm
crit},6}$. This happens at the angular locations $\thetacusp =
\ffrac{\pi}{4} (2n+1)$, where $n$ is an integer.  If the distortion
$\nweb$ increases still further, then the swallowtails can
overlap. This is when eightfold imaging becomes possible.  The
critical value $\nweb_{{\rm crit},8}$ can be determined by requiring
that the two inner pair of cusps just touch each other.  Let us label
the cups in Figure~\ref{fig:cuspthree} in numerical order
counterclockwise starting with 1 on the extreme right hand side.
Then, the condition can be expressed as $\xi(\theta_{\rm cusp,2}) =
\xi(\theta_{\rm cusp,7})$ and $\eta(\theta_{\rm cusp,2}) =
\eta(\theta_{\rm cusp,7})$.  Simultaneously, the 6th and 8th cusps
touch each other as well.  The onset of octuple imaging can be
calculated analytically, but it is rather cumbersome and so we resort
to the computer.  

Figure~\ref{fig:onsetb} shows the domains in which sextuple and
octuple imaging is possible as a function of the ellipticity
$\epsilon$ and the Fourier coefficient $b_4$. Unfortunately, tables of
data of $b_4$ are not available, although $b_4$ is typically smaller
than $a_4$ and rarely exceeds $0.05$. On comparing
Figure~\ref{fig:onsetb} to the earlier Figures~\ref{fig:onset} and
~\ref{fig:onset8}, we see that the skewed squarishness is less
efficient at causing higher order imaging than diskiness or boxiness.

\section{Conclusions}

Many sextuple or octuple images of multiply lensed quasars are likely to
exist.  If a galaxy with a flat rotation curve has exactly elliptic
isophotes, then it yields at most 4 or 5 images (depending on the
behaviour of the convergence or surface density at the centre). Higher
order imaging is caused by deviations of the isophotes from pure
elliptic form, such as boxiness or diskiness.  Even isolated boxy or
disky elliptical galaxies provide caustic networks with swallowtail
and butterfly cusps and so permit sixfold or eightfold imaging if the
source is located propitiously. The rounder the lensing galaxy, then
the smaller the distortion of the isophotes required. The criteria for
the onset of sextuple and octuple imaging can be given exactly in
terms of the ellipticity $\epsilon$ and the Fourier coefficients $a_4$
and $b_4$ used by observers to measure deviations such as boxiness,
diskiness and skewed squarishness (see
eqs.~\ref{eq:cruxa},~\ref{eq:cruxb},~\ref{eq:cruxc}).  

The first examples of galaxy models that can produce sextuple or
octuple images were found by Witt \& Mao (2000) and Keeton et
al. (2000). These models have elliptical density contours with the
external shear perpendicular to the flattening.  These are rare cases,
as they require large values of the external shear for the
cross-section to be substantial -- larger than typically provided by
the effects of large-scale structure. So, one must be lucky to observe
higher order imaging produced by such configurations.  Six and eight
image configurations can also occur when multiple galaxies lie at the
same redshift (Keeton et al. 2000) or at distinct redshifts (Chae, Mao
\& Augusto 2001).  This may be quite common, as the lens in the 6
image case found by Rusin et al (2001) seems to be a cluster. However,
it is perhaps less surprising that groups of galaxies can produce
unusual image configurations.

A crude estimate suggests that about $3 \%$ of the lensing galaxy
population has the potential to produce sextuplets or octuplets.
Unfortunately, the cross-sectional area in the source plane producing
sextuple imaging is approximately a factor of 10 times smaller than
that producing quadruple.  However, the probability of observing a
sextuplet is increased by a factor of $\sim 5$ by the magnification
bias (see e.g. Turner 1980; Turner, Ostriker \& Gott 1984). Therefore,
a realistic estimate is that sextuple and octuple images comprise
$\sim 1 \%$ of all multiply imaged systems. Note that we have ignored
galaxies that show isophote twisting (e.g., Bertola 1981) and galaxies
which exhibit boxiness and diskiness in the same isophotal contours
(e.g., Nieto \& Bender 1989). We have also not included the
contribution from multiple lensing by groups of galaxies. These might
be expected to augment the numbers of sextuplets and octuplets, and so
we reckon our estimate errs on the side of pessimism.

It is reasonable to expect that both the Next Generation Space
Telescope (NGST; see e.g. Barkana \& Loeb 2000) and the Global
Astrometric Interferometer for Astrophysics (GAIA; see e.g., de Boer
et al. 2000) will discover such configurations. For example, GAIA is
the ESA satellite now selected as a Cornerstone 6 mission as part of
the Science
Program~\footnote{http://astro.estec.esa.nl/SA-general/Projects/GAIA/
gaia.html}. It is an all-sky survey satellite that provides
multi-colour, multi-epoch photometry, astrometry and radial velocities
on all objects brighter then $V \approx 21$ (e.g., de Boer et
al. 2000).  It will be sensitive to multiply-imaged quasars with
separations as small as $\sim 0.2^{\prime\prime}$. It is crudely
estimated that GAIA will detect $\sim 4000$ multiply-imaged quasars.
In a dataset of such size, we expect $\sim 40$ sextuple or octuple
image systems.  Both NGST and GAIA will also find ``mini'' Einstein
rings.  We expect that, in many cases, some of the six or eight images
will be merged to form arcs.  On account of the magnification bias, we
expect the lenses to be typically high redshift galaxies and so the
sources themselves will have redshift $\gta 2$.  Therefore, GAIA should
also find ``mini'' Einstein rings, where the diameter of the ring is
sub-arcsecond.

It is interesting to note that any additional cusps produced by
distortion or deviation of elliptical symmetry increases the
magnification of the quadruple image configurations as well. It is
still a puzzle why we observe so many quadruplets. Simulations
typically predict many more double than quadruple image systems.  Boxy
or disky elliptical galaxies favour the occurrence of more quadruplets
This is because the magnification of the 4 images is increased by the
appearance of additional cusps inside the 4 image r\'egime. This fact
has been neglected as only highly symmetric galaxy models have been
typically used in numerical simulations.

The boxy and disky galaxy models with flat rotation curves introduced
in this paper are a natural extension of the popular and widely-used
elliptic density models.  They have many attractive properties.  The
critical curves and the relative time delays are analytic.  The
deflection angles and the caustics are simple quadratures and so
easily evaluated. The lens equation can be reduced to a
one-dimensional equation, and so the image positions can be found
numerically without recourse to grid searching algorithms.  Boxy and
disky galaxy models provide new opportunities to remodel observed
quadruple lenses, where existing models have proved unable to
calculate the flux ratio of the images correctly.  A good candidate
might be MG0414+0534 (see e.g. Hewitt et al. 1992).  Here, we have 4
radio images with well-determined flux ratios; however, no model has
so far been able to reproduce the flux ratios satisfactorily (Falco,
Lehar \& Shapiro 1997). The most recent attempt (Trotter, Winn \&
Hewitt 2000) concludes that the lensing galaxy is somewhat
quadrangular and boxy.

\section*{Acknowledgments}
We thank Shude Mao and Stephen Warren for a number of interesting
discussions and Jeremy Heyl for kindly providing us with data from his
simulations. NWE is supported by the Royal Society. HJW acknowledges
the hospitality of the sub-Department of Theoretical Physics during
working visits.

\end{document}